\documentclass{aa}
\usepackage{graphicx}
\begin{document}

\title{New infrared star clusters in the Northern and Equatorial Milky Way with 2MASS}

\author{ E. Bica \inst{1}, C.M. Dutra \inst{2}, J. Soares \inst{1} and B. Barbuy \inst{2} }

\offprints{C.M. Dutra - dutra@astro.iag.usp.br}

\institute{Universidade Federal do Rio Grande do Sul, Instituto de F\'\i sica, CP\, 15051, Porto Alegre, RS, 91501-970, Brazil\\
\mail{}
\and
 Universidade de S\~ao Paulo, Instituto de Astronomia, Geof\'\i sica e Ci\^encias Atmosf\'ericas, Rua do Mat\~ao 1226, Cid. Universit\'aria,  S\~ao Paulo, SP, 05508-900, Brazil\\
 \mail{}
 }

\date{Received --; accepted --}

\abstract{ We carried out a survey of infrared star clusters and stellar groups on the 2MASS J, H and K$_s$ all-sky release    Atlas in the
Northern and Equatorial Milky Way    (350$^{\circ}$ $< \ell <$ 360$^{\circ}$, 0$^{\circ}$ $< \ell <$ 230$^{\circ}$). The
search in this  zone complements  that in the Southern Milky Way (Dutra et al. 2003a). The method concentrates
efforts on the directions of known optical and radio nebulae. The present study provides 167 new infrared clusters, stellar groups and candidates. Combining the two studies for the whole Milky Way,  346   infrared clusters, stellar groups and candidates were discovered, whereas 315 objects were previously known. They constitute an important new sample for future detailed studies.
\keywords{(Galaxy:) open clusters and associations: general - Infrared:
general}}

\titlerunning{New infrared star clusters in the Northern and Equatorial Milky Way with 2MASS}
\authorrunning{E. Bica et al.}

\maketitle

\section{Introduction}

Embedded  star clusters allow one to study the very initial phases of star formation. Since they are
in general deeply embedded in dust and/or located in heavily reddened lines of sight, the infrared domain
is necessary to study them (e.g. Lada \& Lada 1991; Hodapp 1994;  Deharveng et al. 1997; Carpenter 2000). 

 A large sample of embedded clusters and the knowledge of their 
distribution throughout the Galaxy -- both angularly and in depth -- are fundamental to be established,
which in turn is important for subsequent detailed studies  of individual objects and of the Galactic 
structures to which they belong. Besides serendipitous discoveries and findings in specific directions such as those of 
molecular clouds (Hodapp 1994) or nebulae (Dutra et al. 2003a, hereafter Paper I), systematic methods can be applied:
visual inspections of a whole area (Dutra \& Bica 2000a) or
automated methods (Reyl\'e \& Robin 2002, Ivanov et al. 2002). 

The near infrared  Two Micron All Sky Survey (hereafter 2MASS, 
Skrutskie et al. 1997) has become a fundamental  tool for the discovery of  star clusters and stellar groups 
in the Galaxy, most of them embedded in dust. A literature  compilation of 276 
infrared clusters and stellar groups
(Bica et al. 2003)  included objects reported until mid 2002. Several entries in that catalogue
had been found on the basis
of 2MASS material. In addition to those objects, Dutra \& Bica (2000a) surveyed 2MASS images of central
parts of the Galaxy   and reported 58 small angular size cluster candidates resembling the Arches and Quintuplet
clusters as seen on  2MASS images. Recently, Dutra et al. (2003b) have observed  most of them deeper and at higher angular 
resolution with the ESO NTT telescope: 31 turned out to be blended star images in the  2MASS Atlas, while 27
were confirmed as clusters or remain as cluster candidates.

     More recent discoveries  
are 10  objects from Le Duigou \& Kn\"odlseder (2002) and Ivanov et al. (2002) using 2MASS, 
2 objects from  Deharveng et al. (2002) using  their own observations, and finally, 179 objects  
from  Paper I  using J, H and K$_s$
images from the 2MASS all-sky release  Atlas. 
In Paper I we performed a  search for clusters and stellar groups in the directions of 
known optical and radio nebulae throughout the Southern Milky Way (230$^{\circ}$ $< \ell <$ 350$^{\circ}$).

The present study aims to complete  the search for infrared stellar clusters 
and stellar groups around the Milky Way disk in the directions of known optical and radio nebulae, which was initiated in Paper I. In Sect. 2 we present the search method. In Sect. 3 we provide the newly found objects related to optical
and radio nebulae. In Sect. 4 we discuss some properties of the new samples. Finally, in Sect. 5 concluding remarks are given.

\section{Search method}

Using the same procedures as in Paper I, we searched
for infrared clusters and stellar groups in the J, H and K$_s$ 2MASS images around the central 
positions of optical and  radio nebulae in the Milky Way region 350$^{\circ}$ $< \ell <$ 360$^{\circ}$, 0$^{\circ}$ $< \ell <$ 230$^{\circ}$. We extracted J, H and K$_s$ images 
with 5$^{\prime}\times$5$^{\prime}$ centred on the coordinates
of each nebula from the 2MASS Survey Visualization \& Image Server facility in the web interface {\rm {\it http://irsa.ipac.caltech.edu/}}. For the nebulae with sizes larger than 5$^{\prime}\times$ 5$^{\prime}$ we took additional images with size
10$^{\prime}\times$10$^{\prime}$ or 15$^{\prime}\times$15$^{\prime}$. The K$_s$ band images allow one to probe  deeper
in more absorbed regions, whereas the J and H band images were used to detect blended images by bright star contamination.       

The optical nebula list  was compiled from:
Ber (Bernes 1977), BFS (Blitz et al. 1982), BRC (Sugitani et al. 1991),
Ced (Cederblad 1946), DG
(von Dorschner \& Gurtler 1963), GGD (Gyulbudaghian et al. 1978), GM1- (Gyulbudaghian \& Maghakian 1977a), GM2-
(Gyulbudaghian \& Maghakian 1977b), GM3- (Gyulbudaghian \& Maghakian 1977c), Gy1- (Gyulbudaghian 1982a), Gy2- (Gyulbudaghian 1984a), Gy3- (Gyulbudaghian 1984b), Gy82- (Gyulbudaghian 1982b), NS (Neckel \& Staude 1984), 
Parsamian (Parsamian 1965), PP (Parsamian \& Petrosian 1979), RNO (Cohen 1980), Sh1- (Sharpless 1953), Sh2- (Sharpless 1959) and  vdB-RN (van de Bergh 1966).

Several southern catalogues have extensions to equatorial zones, and  have also been investigated:
ESO (Lauberts 1982), Gum (Gum 1955), RCW (Rodgers et al. 1960) and vdBH-RN (van den Bergh \& Herbst 1975).

The  radio nebula list was compiled   from: G - Reifenstad III et al. (1970), Downes et al. (1980), 
Lockman (1989) and Kuchar \& Clark (1997), CTB - Wilson \& Bolton 
(1960) and Wilson (1963), and W - Westerhout
(1958). We also indicate some infrared nebulae related to sources 
in the AFGL and IRAS catalogues. The list of SNRs is from Green (2001).
The identification of optical SNRs is according to van den Bergh (1978).

Each detection before becoming a discovery was checked against the known objects, i.e.
the infrared catalogue (Bica et al. 2003) and  optical open cluster catalogues  (Alter et al. 1970; 
Lyng\aa~1987;  Dias et al. 2002) in the studied region. 

For the resulting IR star clusters we determined  accurate positions and dimensions from their $K_s$ images (in FITS format) using {\bf SAOIMAGE 1.27.2} developed by Doug  Mink.

\section{Newly found objects}

\begin{figure} 
\resizebox{\hsize}{!}{\includegraphics{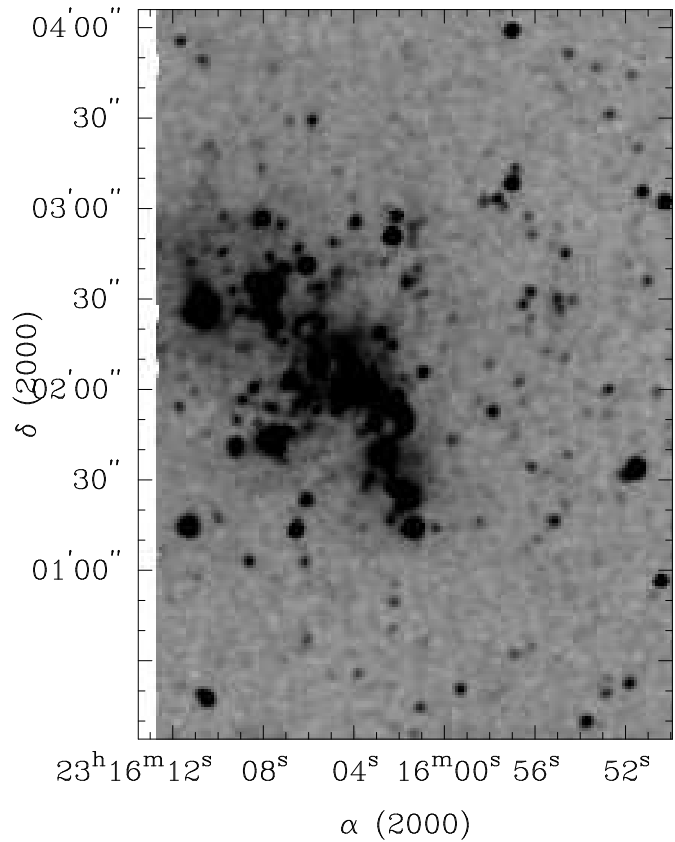}}
\caption[]{4$^{\prime}$ $\times$ 3$^{\prime}$ 2MASS K$_s$ image of   Object 43 in Sh2-157.}
\label{fig1}
\end{figure}

\begin{figure} 
\resizebox{\hsize}{!}{\includegraphics{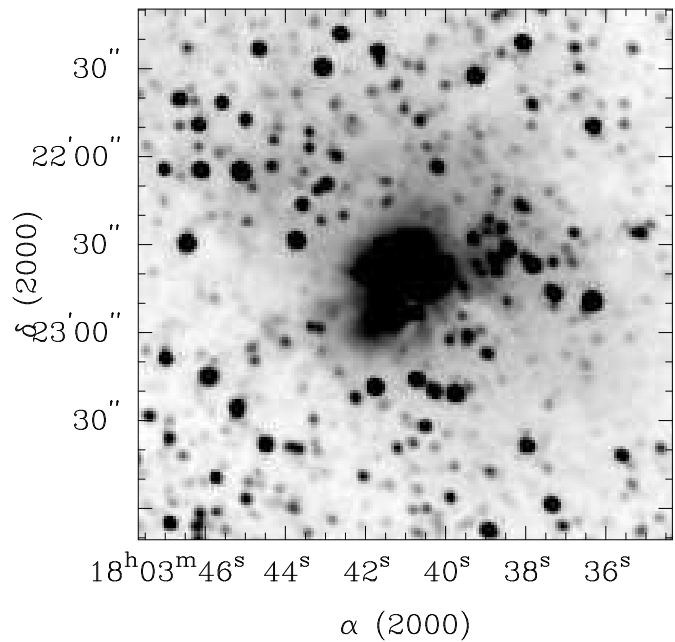}}
\caption[]{3$^{\prime}$ $\times$ 3$^{\prime}$ 2MASS K$_s$ image of   Object 1 in Lagoon Nebula´s Hourglass.}
\label{fig1}
\end{figure}

\begin{figure} 
\resizebox{\hsize}{!}{\includegraphics{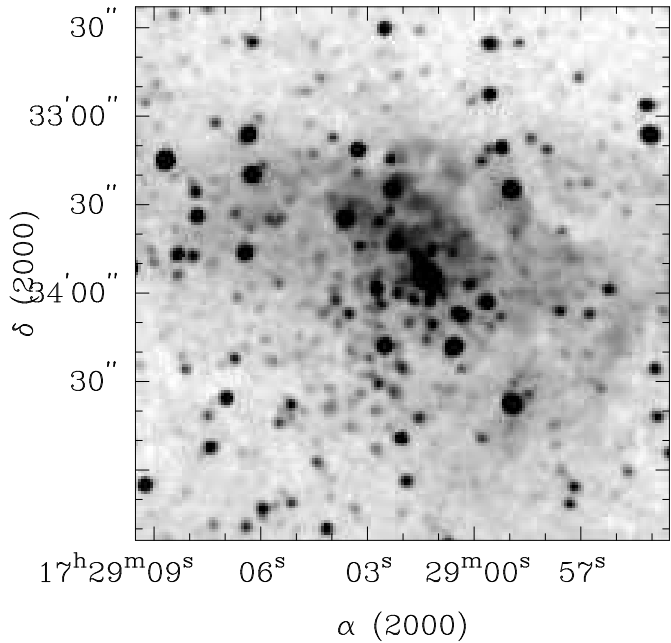}}
\caption[]{3$^{\prime}$ $\times$ 3$^{\prime}$ 2MASS K$_s$ image of Object 166 in the radio nebula G351.694-1.165.}
\label{fig1}
\end{figure}

\begin{figure} 
\resizebox{\hsize}{!}{\includegraphics{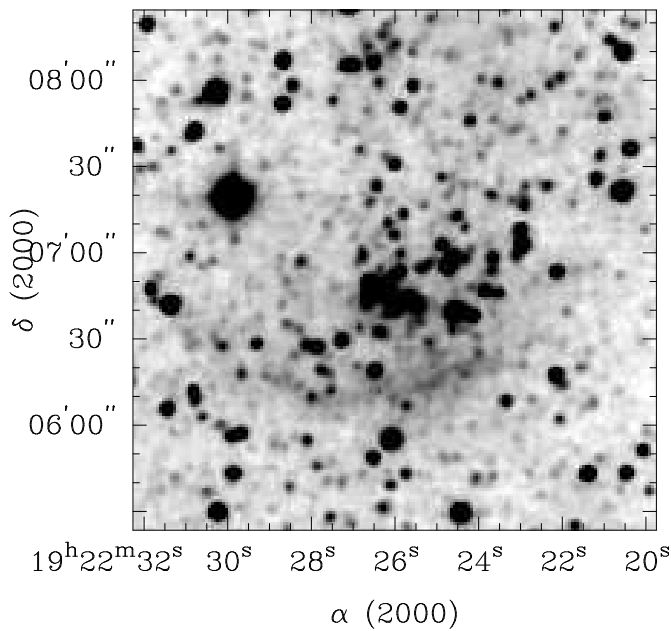}}
\caption[]{3$^{\prime}$ $\times$ 3$^{\prime}$ 2MASS K$_s$ image of Object 139 in the radio nebula W51B=G49.0-0.3.}
\label{fig1}
\end{figure}

Embedded clusters are expected to occur in the areas of nebulae, thus we concentrated our search efforts
on known optical and radio nebulae, mostly HII regions but also reflection nebulae and
supernova remnants.

   We merged the different catalogues and lists of nebulae into a radio/infrared and an 
optical nebula files. We cross-identified
nebulae in each  file, and  between both files. Radio nebulae with optical counterparts were transferred to the
optical nebula file.
The resulting input lists of optical and radio nebulae make in total respectively 1361 and 826 objects in the present Milky Way regions, whose directions were inspected.  The whole Milky Way radio and optical nebula catalogue currently has 4450 entries after cross-identifications, and will be
provided in a forthcoming study. The optical and radio nebula catalogues are similar to the recent dark nebula
catalogue  with 5004 entries by Dutra \& Bica (2002). 

\begin{figure*} 
\resizebox{\hsize}{!}{\includegraphics{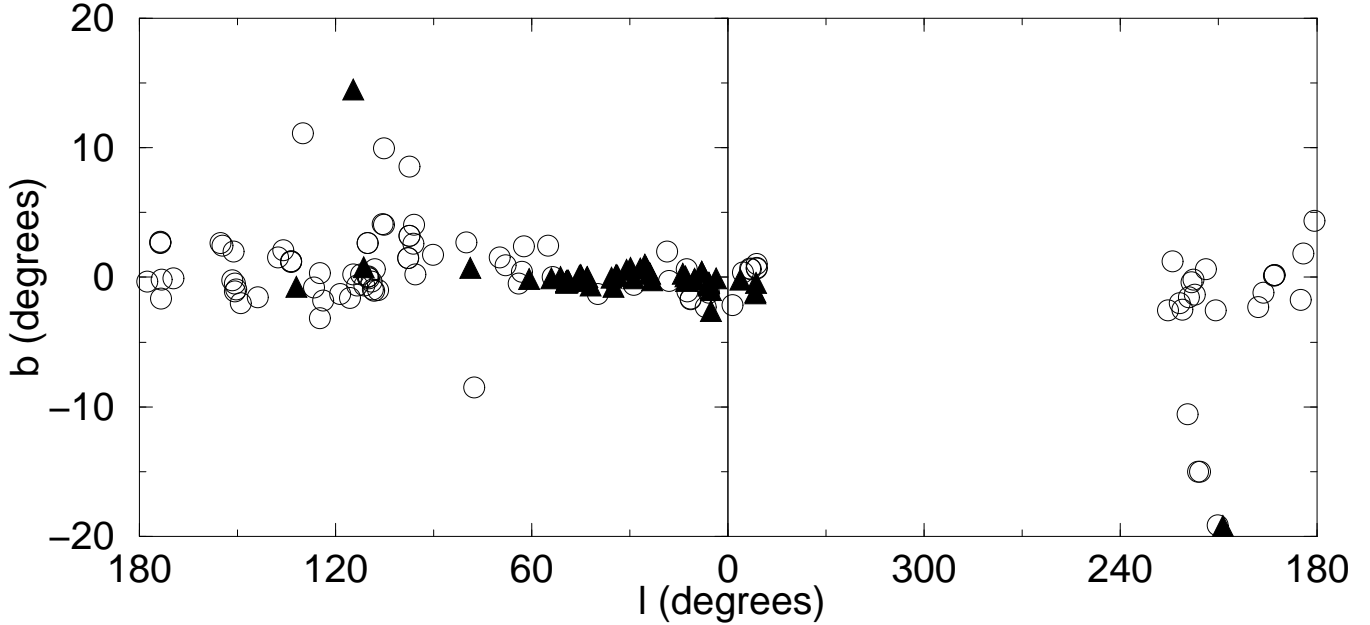}}
\caption[]{Angular distribution of infrared clusters coming  from  optical (Table 1) represented by open circles,  and radio nebulae (Table 2) represented by filled triangles.}
\label{fig1}
\end{figure*}

The results of the cluster survey are shown in Tables 1 and 2, respectively for optical and radio nebulae. 
By Cols.: (1) running number, (2) and (3) Galactic coordinates, (4) and (5) J2000.0 equatorial coordinates, (6)
and (7) major and minor angular dimensions, (8) related nebulae, (9) class, (10) remarks including distance (in case of kinematical 
ambiguity the near and far distances are shown), multiplicity and linear dimension. 

The new infrared clusters, stellar groups and candidates from  the optical nebula survey amount to 103,
and from the radio nebulae survey they are 64. The  detection rates relative to the input nebula catalogues are both 8\% for
optical and radio nebulae. These detection rates are lower than those obtained in Paper I, mostly
because the Equatorial and Northern Milky Way had been previously more surveyed for infrared clusters than their
southern counterpart (Paper I). Several reasons may contribute 
to these low rates: (i) cluster outside the search box owing to projection effects between molecular cloud and 
champagne flow; (ii) extreme absorption in molecular cloud or in the line-of sight; 
(iii) HII region ionized by isolated star or stars; (iv) reflection nebula
illuminated by single star.

We illustrate in Fig. 1 a prominent infrared cluster in an optical nebula: 
Object 43 in Sh2-157.
We show in Fig. 2 a K$_s$ image of Lagoon's Nebula Hourglass (Woodward et al. 1990). With the 2MASS's resolution 
an embedded cluster starts to show up. The western part of the Hourglass is particularly resolved into stars. 
We illustrate in Fig. 3 a prominent infrared cluster located in a radio nebula: 
Object 165 in G351.694-1.165.
In Fig. 4 we show   Object 139 in the radio nebula W51B=G49.0-0.3, part of the large star-forming complex W51.

Object classes are infrared cluster (IRC), stellar group (IRGr), cluster candidate (IRCC),
and open cluster (IROC). Images illustrating  these different classes were given in Paper I.
IRCs are in general populous and at least partially resolved.  
IRCCs are probably clusters, but
are essentially unresolved, and require higher resolution and deeper images for a definitive  diagnostic. 
IRGrs are less dense than IRCs (Bica et al. 2003), some are rather compact but little populated. IROCs 
have similar appearence to optical open clusters and  relatively
large angular size ($\approx$2$^{\prime}$ or more). In Table 1  there occur  45 IRC, 37 IRGr, 14 IRCC and 7 IROC objects,
while in Table 2  16 IRC, 18 IRGr and 30 IRCC objects. The large fraction of cluster candidates among radio 
nebulae certainly reflects higher absorption and/or larger distance effects. 

   Distances are mostly based  on kinematical estimates for the nebulae (Downes et al. 1980), but include as well averages
with estimates from individual stars, when available (e.g. Georgelin et al. 1973). The near/far distance (R)  ambiguity 
has been solved  by Downes et al. (1980) for several nebulae, else we indicate both R$_n$ and R$_f$. 
The provided distances in Tables 1 and 2 should be compared to those 
from the infrared stellar content in future studies.

Some infrared clusters, stellar groups and candidates were found to be related to nebulae which have 
been classified as  planetary nebulae. The present results from inspections of J, H and K$_s$ images 
indicate  or confirm (e.g. Acker 1992) 
that K3-50, Sh2-128, BFS32=PK149-1.1, IC2120=PK169-0.1, Sh2-267=PK196-1.1 and Sh2-271=PK197-2.1 
are not  planetary nebulae.  

\begin{figure} 
\resizebox{\hsize}{!}{\includegraphics{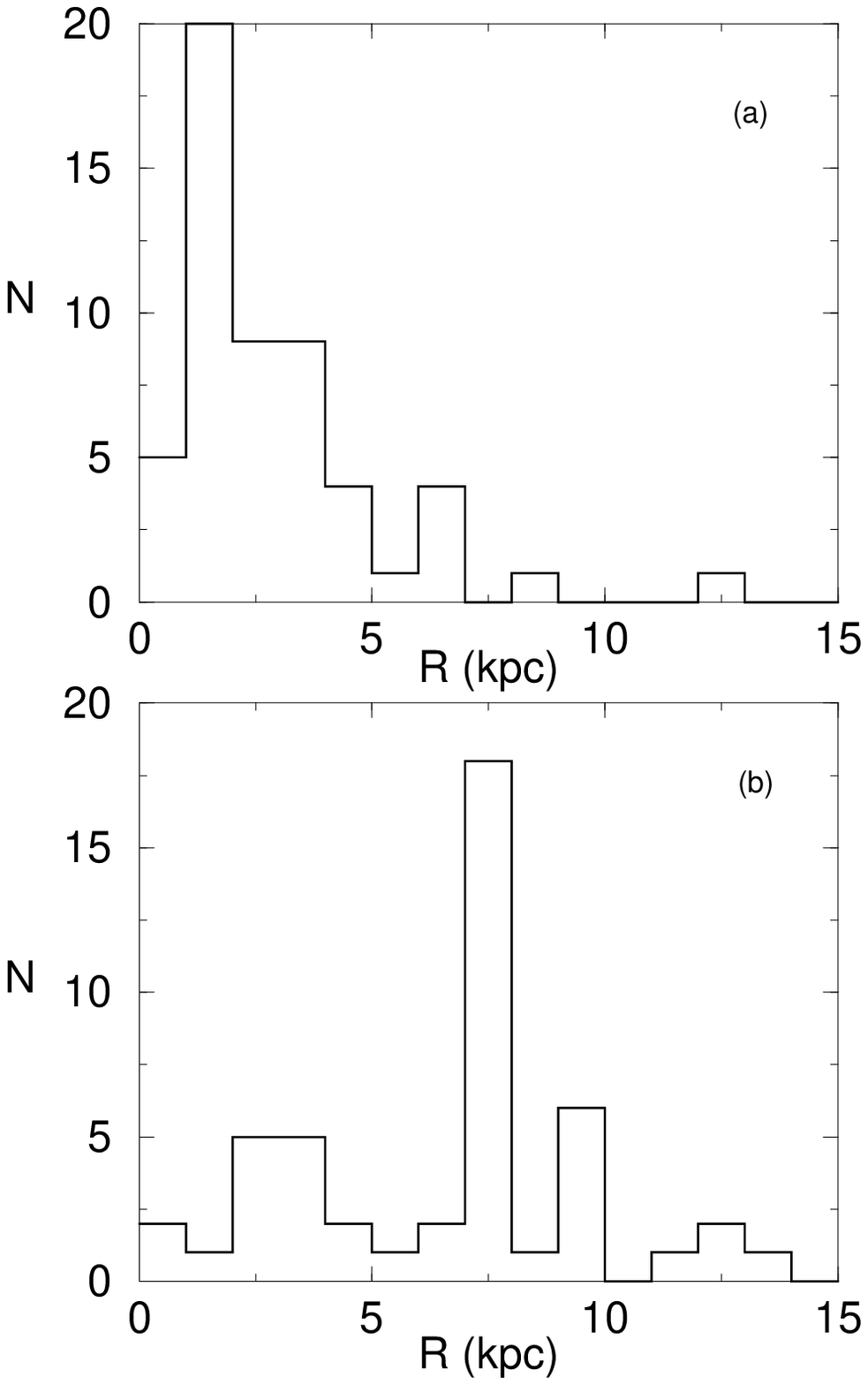}}
\caption[]{Distance histograms for the samples coming from: panel (a)  optical (Table 1) and panel (b) radio nebulae (Table 2).}
\label{fig1}
\end{figure}

\begin{figure} 
\resizebox{\hsize}{!}{\includegraphics{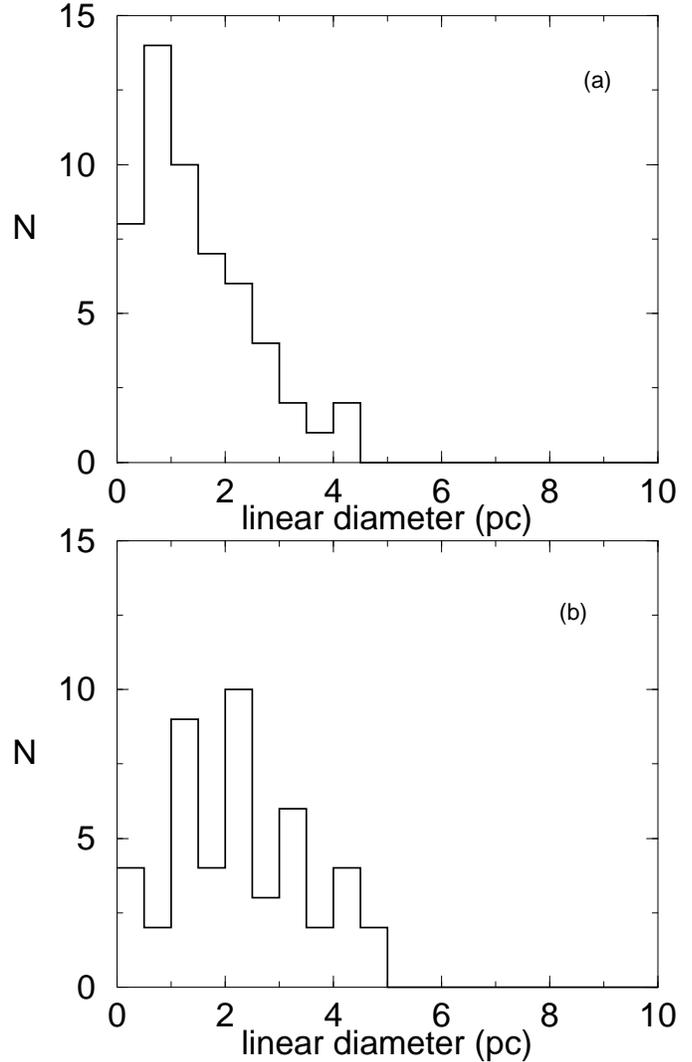}}
\caption[]{Linear major dimension histograms for the samples coming  from: panel (a)  optical (Table 1) and panel (b)  radio nebulae (Table 2).}
\label{fig1}
\end{figure}

\begin{table*}
\caption[]{New objects in the area of optical nebulae}
\begin{scriptsize}
\renewcommand{\tabcolsep}{0.9mm}
\begin{tabular}{lccccccccl}
\hline\hline
Object & $\ell$  &    b& J2000 R.A & J2000 Dec.& D(')& d(') & Related Nebulae& Type& Remarks \\
\hline
  1 &  5.97 & -1.18 &18 03 41 &-24 22 40 &1.9 &1.6 & in Lagoon Neb.=M8=NGC6523=W29=CTB46$^a$       &IRC  &R=1.0 LD=0.6                       \\
  2 &  6.90 & -2.29 &18 09 57 &-24 06 33 &1.9 &1.9 & in NGC6559=ESO521*N40,in Sh2-29               &IRC  &R=1.4 LD=0.8 loose                 \\
  3 &  8.66 & -0.34 &18 06 15 &-21 37 27 &1.1 &1.0 & in Gum77b=RCW151                              &IRCC &R=1.9 LD=0.6                       \\
  4 & 11.48 & -1.65 &18 16 58 &-19 46 58 &2.0 &1.8 & in NGC6589,in Sh1-36=vdB-RN118=ESO590N14      &IRGr &R=1.3 LD=0.8                       \\
  5 & 11.42 & -1.72 &18 17 06 &-19 52 10 &1.6 &1.2 & near IC1283,in RCW153                         &IRGr &                                   \\
  6 & 12.43 & -1.11 &18 16 51 &-18 41 52 &0.9 &0.7 & in Sh2-39                                     &IRCC &                                   \\
  7 & 12.64 &  0.61 &18 10 55 &-17 41 25 &1.9 &1.7 & in vdB-RN116                                  &IRGr &R=0.6 LD=0.3                       \\
  8 & 18.14 & -0.28 &18 25 01 &-13 15 47 &1.8 &1.8 & in G18.143-0.289                              &IRC  &R=4.5 LD=2.4 deeply emb., lane \\
  9 & 18.67 &  1.97 &18 17 53 &-11 44 26 &2.8 &2.3 & in Gum85=G18.7+2.0                            &IRC  &R=2.6 LD=2.1 Opt/IR loose          \\
 10 & 28.97 & -0.60 &18 46 21 & -3 47 42 &1.9 &1.3 & in G28.983-0.603                              &IRGr &                                   \\
 11 & 39.89 & -1.27 &19 08 43 &  5 36 02 &1.4 &1.4 & rel RNO108,in Sh2-74=RCW182=G39.904-1.331     &IRGr &                                   \\
 12 & 53.62 &  0.04 &19 30 23 & 18 20 46 &1.2 &1.2 & in DG159=Ber17,in Sh2-82=AFGL4249             &IRGr &R=1.1 LD=0.4                       \\
 13 & 55.11 &  2.42 &19 24 30 & 20 47 30 &1.3 &1.1 & in Sh2-83=G55.114+2.422                       &IRGr &Opt/IR                             \\
 14 & 62.57 &  2.39 &19 40 22 & 27 18 29 &1.0 &1.0 & in NGC6813=IRAS19383+2711,in W54              &IRC  &                                   \\
 15 & 63.15 &  0.44 &19 49 15 & 26 49 58 &1.9 &1.8 & in Sh2-90=G63.2+0.4                           &IRGr &R=2.0 LD=1.1                       \\
 16 & 64.14 & -0.47 &19 55 00 & 27 12 57 &1.9 &1.7 & in Sh2-93=G64.140-0.470                       &IRC  &R=2.0 LD=1.1                       \\
 17 & 68.14 &  0.92 &19 59 08 & 31 21 38 &1.7 &1.5 & in Sh2-98=G68.134+0.917                       &IRGr &                                   \\
 18 & 69.92 &  1.52 &20 01 10 & 33 11 09 &1.2 &1.2 & in W58G=G69.9+1.5=G69.942+1.517               &IRC  &R=12.4 LD=4.3                      \\
 19$^b$ & 77.69 & -8.48 &21 01 46 & 33 32 34 &0.6 &0.5 & in K3-50=W58A=K3-50A                          &IRCC &R=8.7 LD=1.5 not PN                \\
 20 & 80.03 &  2.69 &20 24 21 & 42 15 56 &1.3 &1.2 & in NGC6914a=DG162=vdB-RN131=Ber23,in Sh2-109  &IRGr &R=1.1 LD=0.4                       \\
 21 & 90.23 &  1.72 &21 05 16 & 49 39 35 &1.8 &1.8 & in Sh2-121=G90.2+1.7                          &IRGr &                                   \\
 22 & 95.67 &  0.24 &21 36 25 & 52 28 04 &1.7 &1.3 & in BFS6=G95.7+0.2                             &IRGr &                                   \\
 23 & 96.03 &  4.03 &21 20 24 & 55 28 01 &2.2 &1.9 & in BFS7                                       &IRGr &Opt/IR                             \\
 24 & 96.29 &  2.59 &21 28 43 & 54 37 03 &1.9 &1.1 & in Sh2-127=G96.3+2.6                          &IRGr &                                   \\
 25 & 97.40 &  8.51 &21 02 49 & 59 30 42 &1.7 &1.5 & in DG168=BFS9=GM1-54=RNO130                   &IRGr &                                   \\
 26 & 97.51 &  3.17 &21 32 10 & 55 52 45 &1.2 &1.2 & in Sh2-128=G97.6+3.2                          &IRCC &mP,not PN                          \\
 27 & 97.53 &  3.18 &21 32 11 & 55 53 42 &0.7 &0.7 & in Sh2-128=G97.6+3.2                          &IRCC &mP,not PN                          \\
 28 & 97.95 &  1.49 &21 42 24 & 54 55 03 &0.8 &0.8 & related to GM1-12=PP101                       &IRCC &mP                                 \\
 29 & 97.97 &  1.49 &21 42 28 & 54 56 00 &1.1 &0.8 & related to GM1-12=PP101                       &IRC  &mP                                 \\
 30 &105.30 &  4.05 &22 14 41 & 61 26 04 &2.1 &2.1 & in DG180                                      &IRC  &loose                              \\
 31 &105.31 &  9.93 &21 42 00 & 66 05 12 &1.9 &1.8 & in GM1-57=PP102=NS20,rel NGC7129              &IRGr &                                   \\
 32 &105.73 &  4.10 &22 17 24 & 61 43 05 &2.3 &1.9 & in RNO140                                     &IRGr &                                   \\
 33 &107.16 & -0.97 &22 47 48 & 58 03 55 &1.7 &1.5 & related to GM2-42,in Sh2-142                  &IRGr &R=3.4 LD=1.7                       \\
 34 &108.16 &  0.60 &22 49 10 & 59 54 54 &1.2 &1.2 & in Sh2-146=G108.197+0.579                     &IRC  &R=4.6 LD=1.6 deeply emb., lane \\
 35 &108.36 & -1.06 &22 56 17 & 58 31 14 &1.7 &1.6 & in Sh2-148                                    &IRC  &R=5.5 LD=2.7 deeply emb.       \\
 36 &108.76 & -0.95 &22 58 41 & 58 46 57 &2.4 &1.8 & in Sh2-152=G108.760-0.952                     &IRC  &R=3.8 LD=2.7 deeply emb.       \\
 37 &109.10 & -0.34 &22 59 06 & 59 28 33 &1.2 &1.0 & related to Gy82-13                            &IRC  &deeply embedded                    \\
 38 &109.99 & -0.09 &23 04 45 & 60 04 36 &1.1 &1.1 & in BFS15                                      &IRC  &R=6.1 LD=2.0                       \\
 39 &110.11 &  0.05 &23 05 11 & 60 14 44 &2.0 &1.5 & in IC1470                                     &IRCC &R=6.1 LD=3.5 deeply emb.       \\
 40 &110.20 &  2.65 &22 56 55 & 62 39 32 &2.9 &1.9 & in Sh2-155=CTB108                             &IRGr &R=0.7 LD=0.6 mP                    \\
 41 &110.21 &  2.62 &22 57 05 & 62 38 16 &1.9 &1.9 & in Sh2-155=CTB108                             &IRC  &R=0.7 LD=0.4 mP loose              \\
 42 &110.25 &  0.01 &23 06 22 & 60 16 11 &1.7 &1.5 & related to BFS18,in Sh2-156=G110.106+0.044    &IRC  &R=6.1 LD=3.0  loose                \\
 43 &111.29 & -0.66 &23 16 06 & 60 02 05 &2.2 &2.0 & in Sh2-157                                    &IRC  &R=2.5 LD=1.6                       \\
 44 &112.22 &  0.22 &23 20 41 & 61 11 35 &1.1 &0.7 & in NGC7635=Sh2-162=G112.237+0.226             &IRGr &R=3.5 LD=1.1 compact,emb.          \\
 45 &113.58 & -0.62 &23 33 34 & 60 50 07 &2.4 &2.1 & in Sh2-163=G113.589-0.721                     &IRC  &R=2.3 LD=1.6 loose                 \\
 46 &114.60 &  0.22 &23 39 44 & 61 55 45 &2.7 &2.1 & in Sh2-165                                    &IRGr &R=1.6 LD=1.3 Opt/IR                \\
 47 &115.78 & -1.58 &23 52 58 & 60 28 30 &2.2 &1.8 & in Sh2-168=G115.784-1.573                     &IRC  &R=3.8 LD=2.4 loose                 \\
 48 &118.62 & -1.32 & 0 15 29 & 61 15 01 &2.8 &2.4 & in Sh2-172                                    &IRC  &loose                              \\
 49 &123.81 & -1.78 & 0 58 40 & 61 04 45 &1.4 &1.4 & in Ced4a=IRAS00556+6046                       &IRGr &                                   \\
 50 &124.87 & -3.14 & 1 06 45 & 59 40 36 &2.3 &1.7 & related to RNO4                               &IRC  &loose                              \\
 51 &124.90 &  0.32 & 1 08 50 & 63 07 40 &1.3 &1.3 & in Sh2-186                                    &IRCC &                                   \\
 52 &126.66 & -0.79 & 1 23 06 & 61 51 23 &2.6 &1.8 & in Sh2-187=Ber49                              &IRC  &Opt/IR                             \\
 53 &130.10 & 11.12 & 2 28 18 & 72 37 48 &1.3 &1.3 & in RNO7                                       &IRC  &loose                              \\
 54$^c$ &133.70 &  1.17 & 2 25 27 & 62 03 33 &1.1 &0.8 & in NGC896,in W3                               &IRC  &R=2.3 LD=0.7 m4                    \\
 55$^c$ &133.69 &  1.22 & 2 25 32 & 62 06 48 &1.2 &0.8 & in W3                                         &IRC  &R=2.3 LD=0.8 m4 deeply emb.        \\
 56$^c$ &133.71 &  1.21 & 2 25 35 & 62 05 36 &1.9 &1.3 & in W3                                         &IRC  &R=2.3 LD=1.3 m4 loose              \\
 57 &136.12 &  2.08 & 2 47 26 & 61 56 53 &1.3 &1.1 & in Sh2-192                                    &IRCC &                                   \\
 58 &137.76 &  1.50 & 2 57 28 & 60 41 37 &2.4 &2.1 & related to LW Cas Nebula=RNO11=PP7$^d$        &IRGr &                                   \\
 59 &143.82 & -1.57 & 3 24 53 & 54 57 25 &1.8 &1.4 & in BFS31                                      &IRGr &                                   \\
 60 &149.08 & -1.99 & 3 51 34 & 51 29 55 &2.7 &1.9 & related to  BFS32=PK149-1.1                   &IROC &not PN                             \\
 61 &150.59 & -0.94 & 4 03 17 & 51 19 35 &3.2 &2.2 & in NGC1491=G150.590-0.950=AFGL5111$^e$        &IRGr &R=3.3 LD=3.1                       \\
 62 &150.86 & -1.12 & 4 03 50 & 51 00 55 &1.3 &1.3 & in Sh2-206=CTB12                              &IRCC &R=3.3 LD=1.2                       \\
 63 &150.98 & -0.47 & 4 07 12 & 51 24 53 &1.2 &1.2 & in BFS34                                      &IRCC &                                   \\
 64 &151.29 &  1.97 & 4 19 33 & 52 58 42 &1.2 &1.2 & in Sh2-208                                    &IRC  &loose                              \\
 65 &151.61 & -0.23 & 4 11 10 & 51 09 58 &2.8 &2.0 & in Sh2-209=G151.594-0.228                     &IRC  &R=4.9 LD=4.0                       \\
 66 &154.65 &  2.44 & 4 36 50 & 50 52 46 &1.6 &1.4 & in Sh2-211=G154.640+2.436                     &IRC  &                                   \\
 67 &155.36 &  2.61 & 4 40 39 & 50 27 39 &2.9 &2.0 & in Ced37=IRAS04366+5022                       &IRC  &loose                              \\
 68 &169.65 & -0.07 & 5 18 11 & 37 33 31 &1.4 &1.4 & in IC2120=PK169-0.1                           &IRGr &not PN                             \\
 69 &173.38 & -0.18 & 5 28 11 & 34 25 28 &3.1 &2.8 & in IC417=Sh2-234                              &IROC &R=2.6 LD=2.3 Opt/IR loose          \\
 70 &173.58 & -1.66 & 5 22 47 & 33 25 26 &2.7 &2.3 & in Ced43=IRAS05189+3327                       &IRC  &loose                              \\
\hline
\end{tabular}
\end{scriptsize}
\begin{list}{}
\item  Notes: $^a $in Hourglass Nebula (Woodward et al. 1990),in Ced152a=IRAS18009-2427=G5.956-1.265, $^b$triplet 
with the NGC6857 and K3-50B clusters (Bica et al. 2003), $^c$the present Objects 54, 55 and 56 form a quadruplet with
the W3-IRS5 Cluster (Bica et al. 2003), $^d$=Gy82-2=IRAS02534+6029, $^e$in Sh2-206, $^f$the present Objects 71, 72
and 73 together with the Sh2-235B Cluster form a quadruplet, $^g$in Ced61=DG86,$^h$in Object NGC6334V (Bica et al. 2003), $^i$in NGC6334=RCW127=CTB39.    
\end{list}
\end{table*}

\begin{table*}
\begin{scriptsize}
\renewcommand{\tabcolsep}{0.9mm}
\begin{tabular}{lccccccccl}
\hline\hline
Object & $\ell$  &    b& J2000 R.A & J2000 Dec.& D(')& d(') & Related Nebulae& Type& Remarks\\
\hline
 71$^f$ &173.76 &  2.66 & 5 40 51 & 35 38 20 &0.8 &0.8 & in GGD5=BFS47=GM2-5=RNO52S,in Sh2-235B        &IRCC &R=1.8 LD=0.4 m4                    \\
 72$^f$ &173.74 &  2.69 & 5 40 54 & 35 40 22 &2.2 &2.0 & in Sh2-235B                                   &IRC  &R=1.8 LD=1.2 m4 loose              \\
 73$^f$ &173.69 &  2.72 & 5 40 55 & 35 44 08 &0.9 &0.9 & in GGD6=GM2-6=RNO52N,in Sh2-235B              &IRCC &R=1.8 LD=0.5 m4 loose              \\
 74 &177.73 & -0.34 & 5 38 47 & 30 41 18 &1.5 &1.2 & in GM1-40=PP36                                &IRGr &deeply embedded                    \\
 75 &180.71 &  4.33 & 6 04 30 & 30 30 07 &3.9 &2.6 & related to vdB-RN65=PP52=IRAS06013+3030$^g$   &IROC &R=1.2 LD=1.4                       \\
 76 &184.00 &  1.83 & 6 01 55 & 26 24 58 &1.1 &1.1 & in GM1-6=BFS49                                &IRC  &                                   \\
 77 &184.87 & -1.73 & 5 50 14 & 23 52 19 &1.3 &1.3 & in IC2144=BFS50                               &IRCC &deeply embedded                    \\
 78 &192.99 &  0.14 & 6 14 23 & 17 44 37 &0.9 &0.9 & in NGC2195                                    &IRC  &                                   \\
 79 &193.01 &  0.13 & 6 14 23 & 17 43 12 &0.9 &0.7 & related to in GM1-45                          &IRGr &mP                                 \\
 80 &192.99 &  0.15 & 6 14 24 & 17 44 42 &1.4 &1.0 & related to in GM1-45                          &IRC  &mP                                 \\
 81 &196.21 & -1.20 & 6 15 53 & 14 16 08 &2.0 &1.7 & in Sh2-267=PK196-1.1                          &IRC  &R=3.5 LD=2.0 not PN                \\
 82 &197.79 & -2.31 & 6 14 57 & 12 21 03 &2.2 &1.9 & in Sh2-271=PK197-2.1                          &IRC  &R=3.3 LD=2.1 not PN,loose          \\
 83 &210.34 & 19.14 & 5 38 24 & -6 24 05 &3.0 &2.1 & in IC428=Ber120,in LDN1641                    &IRGr &R=0.5 LD=0.4                       \\
 84 &210.79 & -2.55 & 6 38 28 &  0 44 41 &1.5 &1.1 & in Sh2-283                                    &IRGr &                                   \\
 85 &213.84 &  0.62 & 6 55 17 & -0 31 26 &0.8 &0.8 & in Sh2-285                                    &IRGr &R=6.9 LD=1.6                       \\
 86 &215.61 & 15.03 & 6 02 16 & -9 06 47 &0.8 &0.6 & in GGD10=GM2-10                               &IRGr &                                   \\
 87 &216.32 & 15.02 & 6 03 30 & -9 43 50 &2.7 &2.7 & in NGC2149=vdB-RN66=PP51=RNO61                &IROC &R=0.8 LD=0.6                       \\
 88 &217.33 & -1.36 & 6 54 36 & -4 32 04 &2.6 &2.0 & in Sh2-286                                    &IRGr &                                   \\
 89 &217.63 & -0.18 & 6 59 24 & -4 15 54 &2.5 &2.0 & in BFS59                                      &IRGr &R=1.4 LD=1.0                       \\
 90 &218.20 & -0.39 & 6 59 41 & -4 51 44 &1.8 &1.5 & related to Gy3-4=RNO81,in Sh2-287S            &IRC  &R=3.2 LD=1.7 loose                 \\
 91 &219.47 & 10.56 & 6 25 16 &-10 33 12 &3.6 &2.3 & related to RNO71                              &IROC &                                   \\
 92 &219.09 & -1.54 & 6 57 11 & -6 11 04 &3.5 &2.9 & related to Parsamian 16=RNO77                 &IROC &                                   \\
 93 &221.01 & -2.51 & 6 57 15 & -8 19 48 &1.1 &1.1 & in RNO78                                      &IRC  &                                   \\
 94 &221.96 & -1.99 & 7 00 51 & -8 56 33 &1.8 &1.6 & in RNO82                                      &IRC  &loose                              \\
 95 &224.17 &  1.22 & 7 16 33 & -9 25 20 &2.1 &1.8 & in Sh2-294=RCW3                               &IRC  &R=4.6 LD=2.8 loose                 \\
 96 &225.48 & -2.57 & 7 05 18 &-12 19 44 &2.7 &1.8 & in Ced90=Gum3=Sh2-297=RCW1a=vdB-RN94=Ber134   &IROC &R=1.1 LD=0.9                       \\
 97$^h$ &351.17 &  0.68 &17 20 03 &-35 58 18 &1.2 &1.2 & rel. to vdBH-RN85b,in NGC6334=RCW127=CTB39    &IRGr &R=1.6 LD=0.6 mP                    \\
 98$^h$ &351.20 &  0.70 &17 20 03 &-35 55 58 &0.9 &0.7 & rel. to vdBH-RN85b,in NGC6334=RCW127=CTB39    &IRC  &R=1.6 LD=0.4 mP                    \\
 99 &351.27 &  1.01 &17 18 59 &-35 41 48 &1.7 &1.3 & rel. to vdBH-RN85a,in Gum63$^i$               &IRGr &R=1.6 LD=0.8                       \\
100 &353.10 &  0.64 &17 25 33 &-34 24 03 &2.0 &1.2 & in RCW131b=G353.1+0.7,in Sh2-11=RCW131=W22    &IRC  &R=1.7 LD=1.0 mP loose              \\
101 &353.11 &  0.65 &17 25 34 &-34 23 08 &0.8 &0.8 & in RCW131b=G353.1+0.7,in Sh2-11=RCW131=W22    &IRC  &R=1.7 LD=0.4 mP loose              \\
102 &355.46 &  0.38 &17 32 52 &-32 34 33 &1.8 &1.3 & in Gum67=Sh2-12=W23=RCW132                    &IRGr &R=1.4 LD=0.7                       \\
103 &358.57 & -2.16 &17 50 47 &-31 16 34 &1.7 &1.7 & in RCW134=W25                                 &IRC  &R=1.9 LD=0.9                       \\
\hline
\end{tabular}
\end{scriptsize}
\begin{list}{}
\item  Notes: Tab. 1 continued.
\end{list}
\end{table*}

We included in Table 1 new embedded clusters and stellar groups related to optical reflection nebulae.
This type of relation has been discussed in Dutra \& Bica (2001). 
These objects appear to be  less massive clusters or stellar groups
where no ionizing star was formed. The objects in Soares \& Bica (2002) are of this type or close to its limit
towards ionizing stars. The present  objects that are related to  van den Bergh$^{\prime}$s (1966) 
and van den Bergh \& Herbst$^{\prime}$s (1975)
reflection nebulae are more probably of this type.

\begin{table*}
\caption[]{New objects in the area of radio/infrared nebulae}
\begin{scriptsize}
\renewcommand{\tabcolsep}{0.9mm}
\begin{tabular}{lccccccccl}
\hline\hline
Object & $\ell$  &    b& J2000 R.A & J2000 Dec.& D(')& d(') & Related Nebulae& Type& Remarks \\
\hline
104 &  3.64 & -0.10 &17 54 25 &-25 51 36 &1.1 &0.9 & in G3.662-0.113                                &IRGr &      R$_n$=1.5 R$_f$=18.5            \\
105 &  5.20 & -2.60 &18 07 30 &-25 44 30 &1.3 &1.0 & related to radio SNR G5.2-2.6?                 &IRGr &                                      \\
106 &  5.34 & -0.99 &18 01 35 &-24 50 06 &1.6 &1.4 & related to radio SNR G5.321-0.974?             &IRGr &                                      \\
107 &  5.90 & -0.43 &18 00 42 &-24 04 23 &1.2 &0.8 & in W28A2=G5.9-0.4=G5.899-0.427,in W28=RCW145   &IRC  &      R=3.0 LD=1.0 mP                 \\
108 &  5.90 & -0.44 &18 00 43 &-24 04 55 &0.45 &0.45 & in W28A2=G5.9-0.4=G5.899-0.427,in W28=RCW145   &IRGr &      R=3.0 LD=0.4 mP compact         \\
109 &  6.15 & -0.64 &18 02 01 &-23 57 40 &1.6 &1.6 & in G6.1-0.6=G6.083-0.117, in W28=RCW145        &IRGr &      R=2.6 LD=1.2                    \\
110 &  8.03 &  0.41 &18 02 05 &-21 48 12 &1.2 &0.9 & in G8.1+0.2=G8.137+0.228                       &IRCC &      mP                              \\
111 &  8.06 &  0.43 &18 02 06 &-21 46 21 &1.8 &1.2 & in G8.1+0.2=G8.137+0.228                       &IRC  &      mP                              \\
112 & 10.31 & -0.14 &18 08 56 &-20 05 30 &1.4 &1.1 & in W31                                         &IRCC &      R=6.0 LD=2.4 mP                 \\
113 & 10.32 & -0.15 &18 09 00 &-20 04 57 &1.3 &1.2 & in W31                                         &IRC  &      R=6.0 LD=2.3 mP                 \\
114 & 12.91 & -0.26 &18 14 40 &-17 52 07 &1.2 &0.8 & in W33C=G12.9-0.3=G12.909-0.277                &IRCC &      R=4.5 LD=1.6 deeply embedded    \\
115 & 13.18 &  0.05 &18 14 05 &-17 28 40 &1.9 &1.7 & in G13.2+0.0=G13.186+0.045                     &IRGr &      R=5.8 LD=3.2\\
116 & 13.88 &  0.28 &18 14 36 &-16 45 17 &1.2 &0.8 & in G13.875+0.282                               &IRC  &      R$_n$=5.6 R$_f$=13.8,deeply emb. \\
117 & 23.23 & -0.24 &18 34 27 & -9 15 44 &1.2 &1.0 & in G22.8-0.5=G22.760-0.485                     &IRCC &      R=12.5 LD=4.4                   \\
118 & 24.19 &  0.29 &18 34 20 & -8 21 27 &1.3 &1.1 & in G23.538-0.041                               &IRCC &      R=11.6 LD=4.4                   \\
119 & 23.93 &  0.28 &18 33 54 & -8 07 32 &1.0 &0.7 & in G23.7+0.2=G23.706+0.171                     &IRCC &                                      \\
120 & 25.58 &  0.99 &18 34 25 & -7 54 50 &0.6 &0.5 & in G24.0+0.2=G23.956+0.152                     &IRGr &      R=12.0 LD=2.1 compact           \\
121 & 25.00 &  0.77 &18 34 10 & -7 18 01 &1.2 &1.0 & in G24.467+0.489                               &IRC  &      R=9.0 LD=3.1 deeply emb.    \\
122 & 26.84 &  0.65 &18 37 58 & -6 53 00 &2.9 &1.9 & in G25.253-0.150,in W42                        &IRC  &      R=4.3 LD=3.8 loose              \\
123 & 28.82 & -0.09 &18 44 15 & -4 17 55 &0.6 &0.5 & in G28.295-0.377                               &IRCC &      compact,deeply embedded         \\
124 & 29.86 &  0.72 &18 43 16 & -3 35 42 &1.2 &1.2 & in G28.801+0.174                               &IRCC &      R=9.0 LD=3.1                    \\
125 & 31.12 &  0.58 &18 46 04 & -2 39 19 &0.6 &0.4 & in G29.9-0.0=G29.944-0.042                     &IRGr &      R=9.0 LD=1.4 compact            \\
126 & 33.92 &  0.11 &18 52 51 &  0 55 28 &1.1 &1.1 & in G33.914+0.111                               &IRCC &      R=8.3 LD=2.7                    \\
127 & 34.26 &  0.15 &18 53 20 &  1 14 39 &1.4 &1.1 & in G34.3+0.1=G34.254+0.144,in W44              &IRC  &      R=3.7 LD=1.5 mP                 \\
128 & 34.25 &  0.13 &18 53 22 &  1 13 58 &0.8 &0.6 & in G34.3+0.1=G34.254+0.144,in W44              &IRCC &      R=3.7 LD=0.9 mP                 \\
129 & 35.20 & -0.74 &18 58 13 &  1 40 37 &1.3 &1.0 & in G35.2-0.74 IR Neb.,in G35.2-0.74 Molec. Cloud$^d$ &IRCC & R=2.3 LD=1.3                    \\
130 & 35.65 & -0.04 &18 56 32 &  2 24 03 &2.0 &1.7 & in G35.663-0.030                               &IRC  &      R=3.5 LD=2.0 loose              \\
131 & 42.12 & -0.62 &19 10 31 &  7 52 57 &1.7 &1.3 & in G42.108-0.623                               &IRC  &      R$_n$=5.1 R$_f$=9.8             \\
132$^e$ & 43.17 &  0.03 &19 10 11 &  9 07 03 &1.0 &0.9 & in W49A=G43.2+0.0,in W49                       &IRGr &      R=13.9 LD=4.0                   \\
133 & 43.22 & -0.05 &19 10 33 &  9 07 37 &1.1 &1.1 & in G43.231-0.054                               &IRGr &      R$_n$=0.6 R$_f$=13.9            \\
134 & 45.13 &  0.14 &19 13 27 & 10 54 27 &1.2 &1.2 & in G45.125+0.136                               &IRCC &      R=9.7 LD=3.4 mP                 \\
135 & 45.12 &  0.13 &19 13 28 & 10 53 35 &0.7 &0.7 & in G45.125+0.136                               &IRCC &      R=9.7 LD=2.0 mP compact         \\
136 & 45.48 &  0.13 &19 14 09 & 11 12 32 &1.3 &1.0 & in G45.475+0.130                               &IRCC &      R=9.7 LD=3.7 deeply embedded    \\
137 & 45.82 & -0.29 &19 16 19 & 11 19 08 &1.6 &1.4 & in G45.8-0.3=G45.824-0.290                     &IRCC &                                      \\
138 & 48.92 & -0.28 &19 22 15 & 14 03 32 &1.9 &1.6 & in G48.9-0.3=G48.930-0.286,in W51              &IRC  &      R=7.5 LD=4.1 mP deeply emb. \\
139 & 48.99 & -0.30 &19 22 26 & 14 06 54 &2.2 &2.0 & in W51B=G49.0-0.3,in W51                       &IRC  &      R=7.5 LD=4.8 mP loose           \\
140 & 49.06 & -0.28 &19 22 30 & 14 11 03 &2.1 &2.1 & in G49.1-0.3=G49.060-0.260,in W51              &IRCC &      R=7.5 LD=4.6                    \\
141 & 49.08 & -0.37 &19 22 53 & 14 09 22 &1.5 &1.3 & in G49.1-0.4=G49.076-0.377,in W51              &IRGr &      R=7.5 LD=3.3 deeply embedded    \\
142 & 49.38 & -0.27 &19 23 04 & 14 28 05 &1.0 &1.0 & in G49.4-0.3=G49.384-0.298,in W51              &IRCC &      R=7.5 LD=2.2                    \\
143 & 49.39 & -0.30 &19 23 14 & 14 27 33 &1.4 &1.3 & in G49.4-0.5=G49.437-0.465,in W51              &IRGr &      R=7.5 LD=3.1 mP                 \\
144 & 49.42 & -0.31 &19 23 19 & 14 29 23 &1.3 &1.0 & in W51                                         &IRCC &      R=7.5 LD=2.8 mP                 \\
145 & 49.48 & -0.33 &19 23 29 & 14 31 43 &1.3 &1.3 & related to W51A=G49.5-0.4=G49.486-0.381,in W51 &IRCC &      R=7.5 LD=2.8 mP                 \\
146 & 49.49 & -0.34 &19 23 35 & 14 32 02 &0.9 &0.9 & related to W51A=G49.5-0.4=G49.486-0.381,in W51 &IRCC &      R=7.5 LD=2.0 mP   \\
147 & 49.46 & -0.35 &19 23 33 & 14 29 47 &0.6 &0.6 & in W51                                         &IRCC &      R=7.5 LD=1.3                    \\
148 & 49.46 & -0.39 &19 23 41 & 14 29 15 &0.6 &0.6 & in W51A,G49.5-0.4,G49.486-0.381,in W51         &IRCC &      R=7.5 LD=1.3 m5                 \\
149 & 49.48 & -0.39 &19 23 43 & 14 29 55 &1.5 &0.7 & in W51A,G49.5-0.4,G49.486-0.381,in W51,$^c$    &IRCC &      R=7.5 LD=3.3 m5 deeply emb. \\
150 & 49.49 & -0.38 &19 23 42 & 14 30 47 &0.6 &0.3 & in W51A,G49.5-0.4,G49.486-0.381,in W51,$^b$    &IRCC &      R=7.5 LD=1.3 m5 deeply emb. \\
151 & 49.49 & -0.38 &19 23 43 & 14 30 34 &0.8 &0.6 & in W51A,G49.5-0.4,G49.486-0.381,in W51,$^b$    &IRCC &      R=7.5 LD=1.7 m5 deeply emb. \\
152 & 49.49 & -0.37 &19 23 40 & 14 31 13 &0.7 &0.7 & in W51A,G49.5-0.4,G49.486-0.381,in W51,$^a$    &IRCC  &      R=7.5 LD=1.5 m5 deeply emb. \\
153 & 49.54 & -0.38 &19 23 48 & 14 33 15 &1.1 &0.9 & in W51                                         &IRCC &      R=7.5 LD=2.4 mP deeply emb. \\
154 & 49.54 & -0.39 &19 23 51 & 14 32 57 &1.0 &1.0 & in W51                                         &IRCC  &      R=7.5 LD=2.2 mP deeply emb. \\
155 & 49.59 & -0.39 &19 23 55 & 14 35 40 &1.1 &1.1 & in G49.6-0.4=G49.582-0.382,in W51              &IRC  &      R=7.5 LD=2.4 deeply embedded    \\
156 & 54.08 & -0.07 &19 31 43 & 18 41 57 &1.8 &1.8 & in G54.092-0.066                               &IRGr &      R$_n$=3.9 R$_f$=7.9             \\
157 & 51.36 & -0.01 &19 26 02 & 16 20 10 &2.3 &1.7 & in G51.362-0.001                               &IRGr &      R$_n$=5.0 R$_f$=7.5             \\
158 & 60.88 & -0.13 &19 46 20 & 24 35 22 &1.1 &0.9 & in Sh2-87 IR Nebula=G60.888-0.127              &IRC  &      R=2.7 LD=0.9 deeply embedded    \\
159 & 78.88 &  0.71 &20 29 24 & 40 11 14 &0.9 &0.8 & related to AFGL2591$^f$                        &IRCC &      R=1.0 LD=0.3 deeply embedded    \\
160$^g$ &111.56 &  0.75 &23 13 59 & 61 27 01 &1.5 &1.1 & in NGC7538-IRS9                                &IRCC &      R=2.8 LD=1.2 mT deeply emb. \\
161 &114.63 & 14.51 &22 38 45 & 75 11 38 &1.5 &1.0 & in LDN1251B IR Nebula, in dark nebula LDN1251  &IRGr &      R=0.2 LD=0.1                    \\
162 &132.16 & -0.73 & 2 08 05 & 60 45 53 &1.9 &1.5 & in G132.2-0.7=G132.157-0.725                   &IRGr &                                      \\
163$^h$ &208.72 & -19.20 & 5 35 27 & -5 03 56 &1.5 &0.9 & related to IRAS05329-0505, in OMC-3            &IRGr &      R=0.5 LD=0.2                    \\
164 &351.47 & -0.46 &17 25 32 &-36 21 58 &0.8 &0.8 &in G351.467-0.462                               &IRC  &      R$_n$=4.0 R$_f$=15.8            \\
165 &351.64 & -1.25 &17 29 17 &-36 40 03 &2.0 &1.4 &in G351.6-1.3=G351.613-1.270                    &IRC  &      R$_n$=2.6 R$_f$=17.2 deeply emb. \\
166 &351.69 & -1.15 &17 29 02 &-36 33 53 &1.8 &1.5 &in G351.694-1.165                               &IRC  &      R=2.7 LD=1.4                    \\
167 &356.30 & -0.19 &17 37 18 &-32 10 48 &0.7 &0.6 & in G356.307-0.210=IRAS17341-3208               &IRGr &      R$_n$=1.0 R$_f$=19.0            \\
\hline
\end{tabular}
\end{scriptsize}
\begin{list}{}
\item  Notes: $^a$in W51d=W51-IRS2 (Goldader \& Wynn-Wiliams 1994) , $^b$in W51e=W51-IRS1-North (Goldader \& Wynn-Wiliams 1994), $^c$in W51-IRS1-South (Goldader \& Wynn-Wiliams 1994),$^d$ Tapia et al. (1985), $^e$ quadruplet with W49A, 
W49A-east and W49A-southwest clusters (Bica et al. 2003), $^f$the 2MASS images indicate several stellar sources
which suggest a small cluster or stellar group around the massive YSO (Marengo et al. 2000), $^g$ NGC7538 IR Cluster in Bica et al. (2003)
can be divided into a large loose NW and a compact SE  (NGC7538-IRS1 Cluster) components. These two clusters 
together with  the present object form a triplet, $^h$ the 2MASS images indicate  stellar sources
which suggest a small stellar group.    
\end{list}
\end{table*}

\section{Discussion}

\begin{figure*} 
\resizebox{\hsize}{!}{\includegraphics{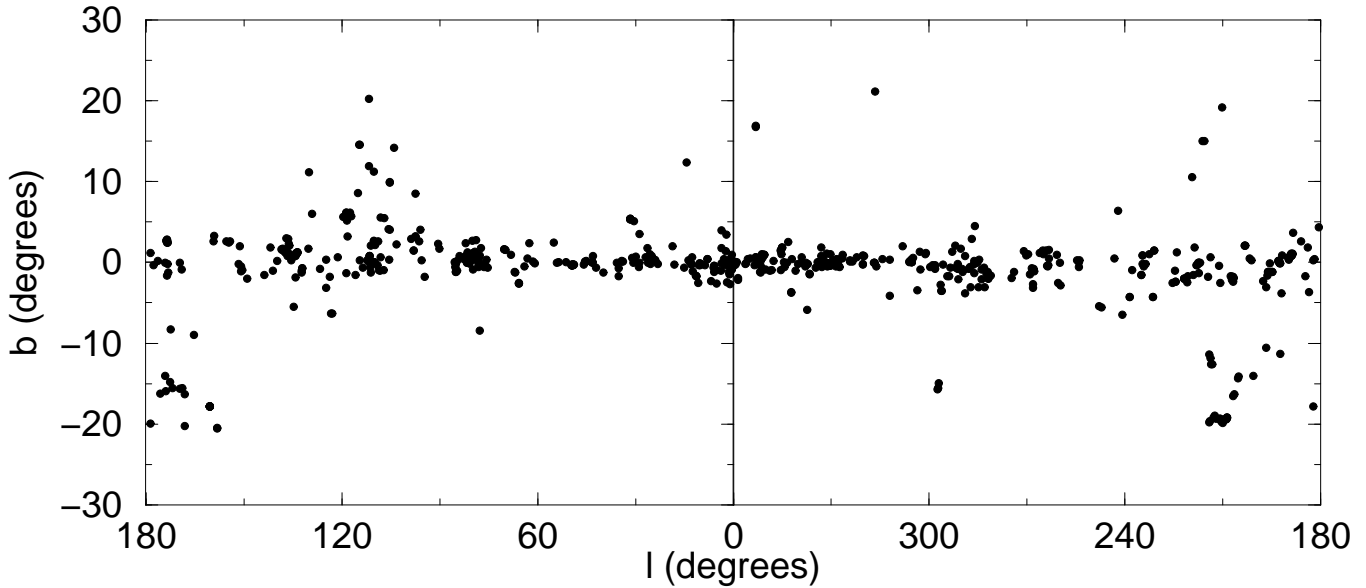}}
\caption[]{Angular distribution of infrared clusters of all known IR clusters, stellar groups and candidates}
\label{fig1}
\end{figure*}

 For the present sample (Tables 1 and 2) there occur 42 clusters 
which are members of pairs, triplets, quadruplets or quintuplet. The quintuplet occurs in W51.
That number  corresponds to a fraction of
25\%, confirming the properties of the two previous samples (Bica et al. 2003 and Paper I). In the present sample multiplicity occurs more often
among objects coming from the radio sample, which amount to 24 objects. As discussed in our two previous papers (Bica et al. 2003 and Paper I) multiplicity
must play a significant  role in the early dynamical evolution of star clusters.

   The present results confirm that W51 is a prominent star forming complex  in the Milky Way (Goldader \& Wynn-Wiliams 1994 and references therein), or that some depth effects occur
since this is the eastern 
tangent point of the Sagittarius-Carina Arm. Table 2 shows that 18 objects are related to W51: 3 IRCs, 2 IRGrs and 13 IRCCs.

  Fig. 5 shows the angular distributions of the present samples (Tables 1 and 2).
Objects from the radio sample  are mainly located between $350^{\circ} < \ell < 360^{\circ}$ and $0^{\circ} < \ell < 60^{\circ}$ corresponding to  internal arms and where absorption in the Galaxy shows a pronounced increase (Dutra \& Bica 2000b). Objects coming from the optical sample are more evenly distributed across the surveyed region, including
directions of the  Local, Perseus and Outer Arms. Both samples have objects in directions of the Sagittarius-Carina
Arm, but more probably objects related to optical nebulae belong to it.

  Fig. 6 shows the distance histograms for the objects (Tables 1 and 2). In cases of  distance ambiguity we assumed their near distance as a lower limit for the histogram analysis. 
The optical sample has a pronounced peak at 2 kpc, but has important contributions up to 4 kpc.
The histogram of the radio  sample covers a wide distance range. The pronounced peak at 7.5 kpc corresponds to the large number of objects in the W51 complex. A secondary peak occurs for $\approx$9 kpc.  Radio
objects are also significantly distributed in the range 2-4 kpc.

  Fig. 7 shows the linear size histograms for the objects (Tables 1 and 2). 
Objects with distance ambiguity were excluded. The histogram of objects coming
from the optical nebulae sample is  remarkably similar to that obtained in
Bica et al. (2003) and its counterpart of Paper I, skewed with a peak at about 1 pc. The histogram of objects 
from the radio nebulae sample is  more evenly distributed, with a peak at about 2 pc, likewise its counterpart
in Paper I. 

Concerning SNRs, one probable IRGr is in the area of the radio SNR
GG5.2-2.6, and one in that of G5.321-0.974 (Table 2), confirming that such relations are a rare phenomenon (Paper I).

\begin{figure} 
\resizebox{\hsize}{!}{\includegraphics{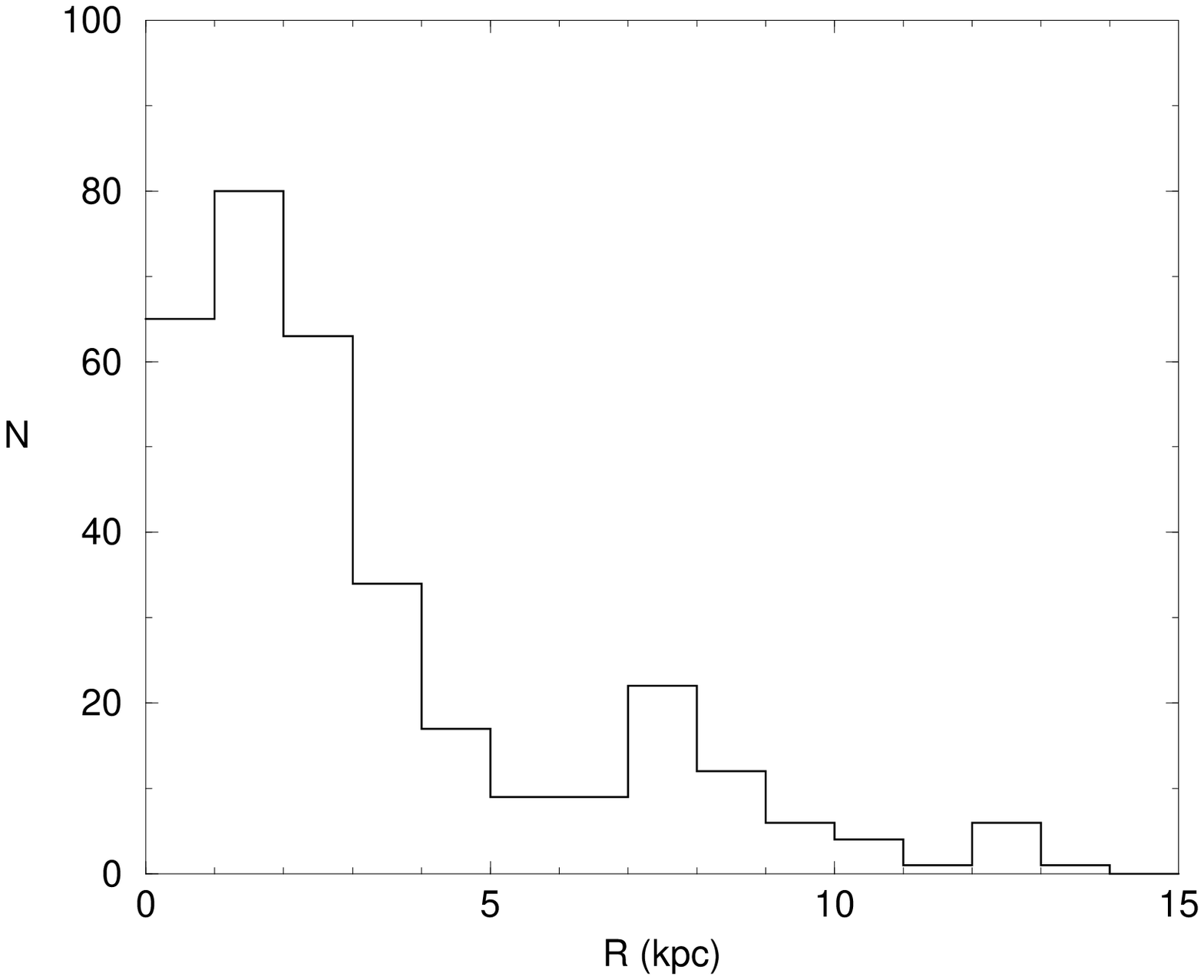}}
\caption[]{Distance histogram for all known IR clusters, stellar groups and candidates}
\label{fig1}
\end{figure}

\begin{figure} 
\resizebox{\hsize}{!}{\includegraphics{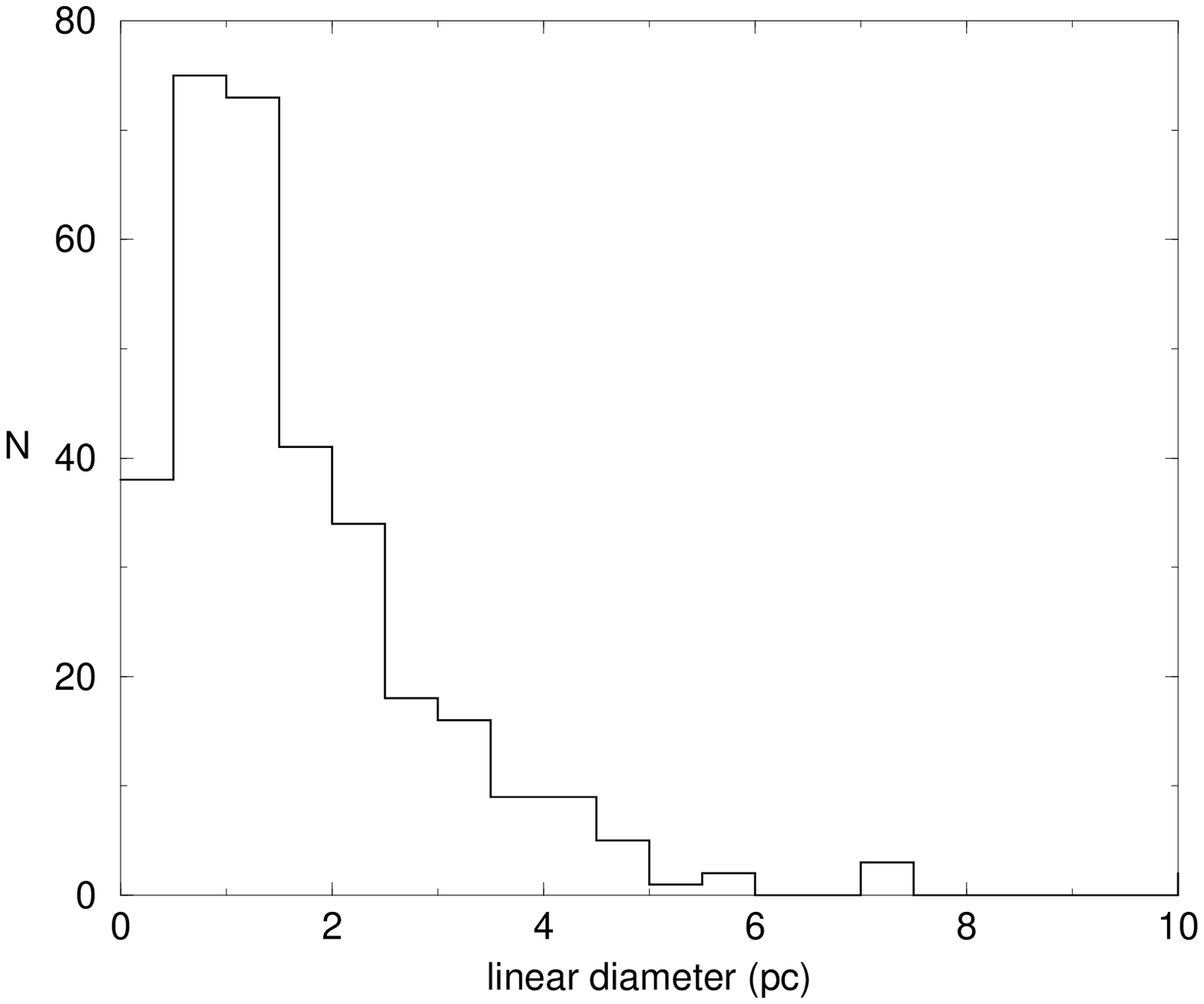}}
\caption[]{Linear major dimension histogram for all known IR clusters, stellar groups and candidates}
\label{fig1}
\end{figure}

\subsection {Total sample of known objects}

Considering the results of the present work, Paper I and the literature objects indicated in Sect. 1, the total
sample of known infrared clusters, stellar groups and candidates is now 661.

Fig. 8  shows the angular distribution in galactic coordinates  of the total sample. Now the whole Milky
Way has been more uniformly surveyed, which was  not the case of the literature sample (see Fig. 1 of Bica et al. 2003).
The objects towards the central parts are distributed more closely  to the plane, suggesting larger distances on the average.
The objects at higher latitudes around the anticentre are in the nearby Orion and Taurus complexes.

Fig. 9 shows the distance histogram for the total sample. The distribution peak occurs at 1.5 kpc, showing that a typical IR cluster or stellar group is a relatively 
close object. Most of the objects are in the range 1-4 kpc, but an important sample also occurs for
5-9 kpc. A few objects exceed 10 kpc.  

Fig. 10 shows the linear size histogram for the total sample. 
The enormous sample that we are dealing with  shows that embedded clusters are typically small, less
than 2 parsecs in size, and hardly exceeding 4 pc. This must reflect the dimensions of the regions 
in molecular clouds where
star formation is efficient enough to create zones of high stellar densities. The present results suggest that 
the extent of such regions does not vary much from cloud to cloud.

We can also estimate the density increase of objects along the Galactic disk by comparing
the samples in Bica et al. (2003) and the present total sample. Considering clusters with $|b|<$ 5$^{\circ}$
the former sample provided 0.53 objects (degree)$^{-1}$ along Galactic longitude, while the present overall sample
yields 1.57 (degree)$^{-1}$.  

The angular distributions and distances indicate  that the now known overall sample of embedded 
clusters and stellar groups arises mostly in external arms, in the near side 
of internal arms and to a certain extent central parts of the Galaxy.
Beyond  these regions, especially towards the inner Galaxy, line-of-sight absorption effects must be too strong, even for the near IR domain.

\subsubsection{Sensitivity of surveys}

It is interesting to compare the spatial distribution of infrared clusters and stellar groups 
with those of the optical and radio nebulae.  A decrease of clusters with distance 
would favour   distance/reddening effects on detectability
while other distributions  might suggest  that many nebulae  do not harbour any cluster. We study the
available objects in the whole plane, including the present sample, that of Paper I and the previously 
catalogued clusters (Bica et al. 2003).  
Distance
information is available for 958 optical and 358 radio nebulae. There are 232 clusters or groups related to
optical nebulae and 119  to radio nebulae with distance estimates.  The comparison is made in Fig. 11. 
The blowup for optical objects (upper right panel) suggests a detection decrease
with distance for d $>$ 2.5 kpc. For the radio sample the distance decrease is clear for d $>$ 5 kpc (lower
left panel). These results suggest that  nebulae with distance information in general harbour a cluster or
stellar group and that many remain undetected owing to a reddening/distance horizon effect. However,
the total number of nebulae in the catalogues is much larger (Sect. 3) than those with velocity (distance) information,
which suggests that many deal with structural details of the nebular complexes, which do not necessarily harbour
any cluster.

\section{Concluding remarks}

We searched for embedded star clusters and stellar groups
in the directions of 1361 optical and 826 radio nebulae in the Equatorial and Northern Milky Way (in the region
350$^{\circ}$ $< \ell <$ 360$^{\circ}$, 0$^{\circ}$ $< \ell <$ 230$^{\circ}$) using J, H and 
K$_s$ images from the 2MASS all-sky
release Atlas. A total of 167 new infrared clusters, stellar groups and candidates were found. 
Together with 179 discoveries from  Paper I, the present method provided a
total of 346 new objects. This number is larger than that of  all previously known infrared clusters, stellar groups and
candidates in the literature, which amount to 315 objects. 

The physical properties of the present sample are similar to those of its southern counterpart (Paper I), in particular
concerning the size distribution of clusters coming from the optical and radio nebulae samples. 
Multiplicity appears to affect about 25\% of the embedded clusters,
suggesting that interactions and mergers can affect their early dynamical evolution.
Objects from the optical nebulae sample are on the average closer (at 2-4 kpc) than those from the radio nebulae 
sample. 

The present contribution and that of Paper I provide a fundamental new sample for
detailed future studies. Resolved brighter  objects may be studied with 2MASS photometry itself, but
most of the sample requires large telescopes for deep photometry. 

Considering the results of the present work, Paper I and the literature objects indicated in Sect. 1, the total
sample of known infrared clusters, stellar groups and candidates now  amounts to 661 objects. 
Most are located nearby  (1-4 kpc) and they are typically smaller than 2 pc. 

Much work is yet to be done 
with 2MASS, especially in the search of more evolved disk clusters away from nebulae. Systematic visual
searches in specific areas and searches using automated methods will certainly provide many new 
discoveries.
 
 \begin{figure*} 
\resizebox{\hsize}{!}{\includegraphics{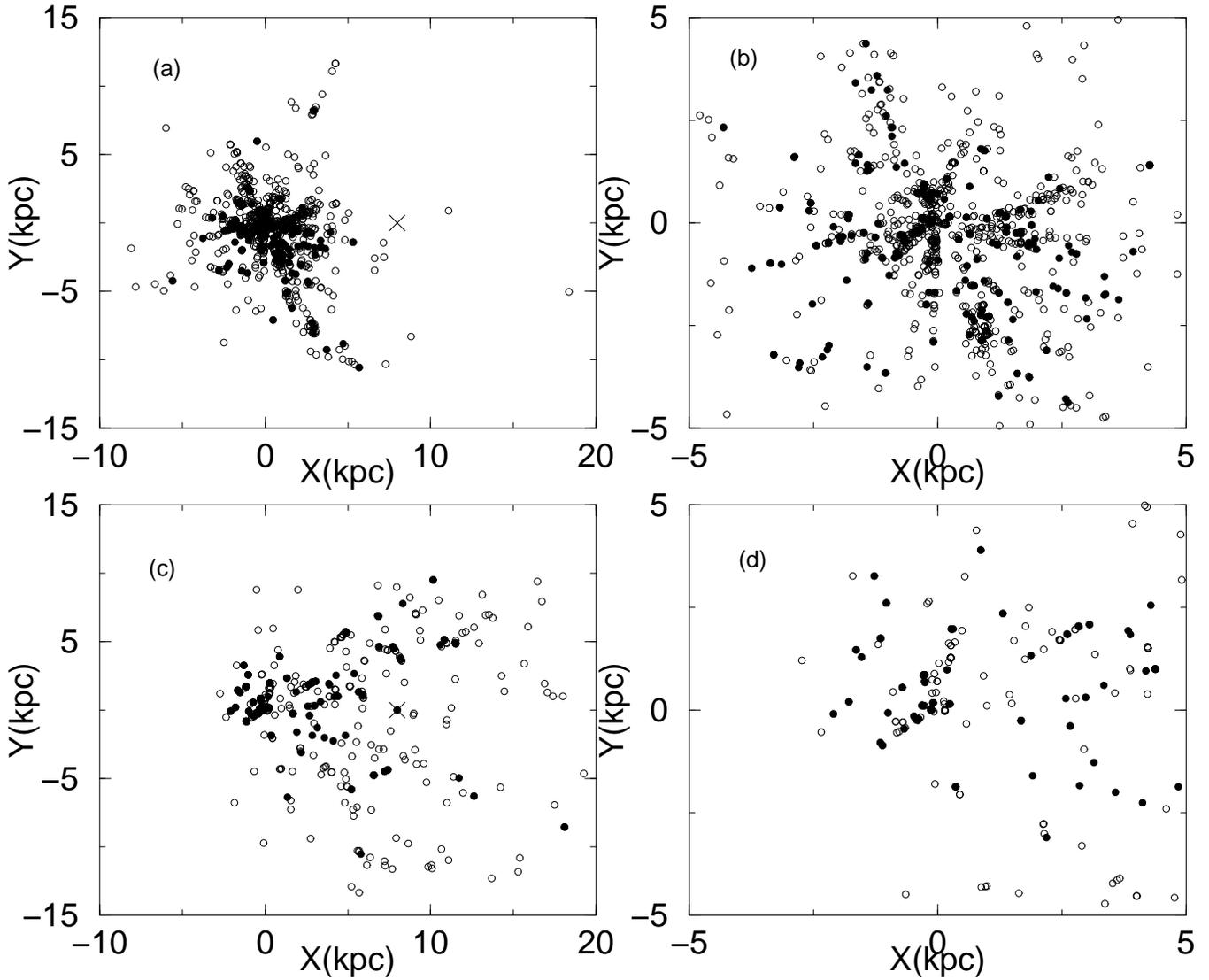}}
\caption[]{Projection on the Galactic plane (heliocentric coordinates) of embedded clusters and nebulae.
Upper left and lower left panels correspond to optical and radio nebulae, respectively. Upper right and
lower right panels are the respective blowups. Open circles are nebulae, filled circles are clusters. Cross
indicates the Galactic centre.}
\label{fig1}
\end{figure*}
 
\begin{acknowledgements}
This publication makes use of data products from the Two Micron All Sky Survey, which is a joint project of the University of Massachusetts and the Infrared Processing and Analysis Center/California Institute of Technology, funded by the National Aeronautics and Space Administration and the National Science Foundation.
We employed  data from CDS/Simbad (Strasbourg). We thank the referee Dr. F. Comer\'on for
interesting remarks. We acknowledge support from the Brazilian Institutions CNPq and FAPESP. CMD acknowledges FAPESP for a post-doc fellowship (proc. 00/11864-6).
\end{acknowledgements}

%
%

\end{document}